\documentclass[12pt]{article}

\usepackage{amsmath,amsthm,latexsym,amssymb,amsfonts}
\usepackage{latexsym}

\makeatletter
\@addtoreset{equation}{section}
\makeatother
\newtheorem{Theorem}{Theorem}[section]
\newtheorem{Definition}{Definition}[section]

\def\be{\begin{equation}}
\def\ee{\end{equation}}
\def\ba{\begin{eqnarray}}
\def\ea{\end{eqnarray}}
\def\Nl{{\mathchoice
{\setbox0=\hbox{$\displaystyle\rm N$}\hbox{\hbox to0pt
{\kern0.4\wd0\vrule height0.9\ht0\hss}\box0}}
{\setbox0=\hbox{$\textstyle\rm N$}\hbox{\hbox to0pt
{\kern0.4\wd0\vrule height0.9\ht0\hss}\box0}}
{\setbox0=\hbox{$\scriptstyle\rm N$}\hbox{\hbox to0pt
{\kern0.4\wd0\vrule height0.9\ht0\hss}\box0}}
{\setbox0=\hbox{$\scriptscriptstyle\rm N$}\hbox{\hbox to0pt
{\kern0.4\wd0\vrule height0.9\ht0\hss}\box0}}}}
\def\Zl{{\mathchoice
{\setbox0=\hbox{$\displaystyle\rm Z$}\hbox{\hbox to0pt
{\kern0.4\wd0\vrule height0.9\ht0\hss}\box0}}
{\setbox0=\hbox{$\textstyle\rm Z$}\hbox{\hbox to0pt
{\kern0.4\wd0\vrule height0.9\ht0\hss}\box0}}
{\setbox0=\hbox{$\scriptstyle\rm Z$}\hbox{\hbox to0pt
{\kern0.4\wd0\vrule height0.9\ht0\hss}\box0}}
{\setbox0=\hbox{$\scriptscriptstyle\rm Z$}\hbox{\hbox to0pt
{\kern0.4\wd0\vrule height0.9\ht0\hss}\box0}}}}
\def\Ql{{\mathchoice
{\setbox0=\hbox{$\displaystyle\rm Q$}\hbox{\hbox to0pt
{\kern0.4\wd0\vrule height0.9\ht0\hss}\box0}}
{\setbox0=\hbox{$\textstyle\rm Q$}\hbox{\hbox to0pt
{\kern0.4\wd0\vrule height0.9\ht0\hss}\box0}}
{\setbox0=\hbox{$\scriptstyle\rm Q$}\hbox{\hbox to0pt
{\kern0.4\wd0\vrule height0.9\ht0\hss}\box0}}
{\setbox0=\hbox{$\scriptscriptstyle\rm Q$}\hbox{\hbox to0pt
{\kern0.4\wd0\vrule height0.9\ht0\hss}\box0}}}}
\def\Rl{{\mathchoice
{\setbox0=\hbox{$\displaystyle\rm R$}\hbox{\hbox to0pt
{\kern0.4\wd0\vrule height0.9\ht0\hss}\box0}}
{\setbox0=\hbox{$\textstyle\rm R$}\hbox{\hbox to0pt
{\kern0.4\wd0\vrule height0.9\ht0\hss}\box0}}
{\setbox0=\hbox{$\scriptstyle\rm R$}\hbox{\hbox to0pt
{\kern0.4\wd0\vrule height0.9\ht0\hss}\box0}}
{\setbox0=\hbox{$\scriptscriptstyle\rm R$}\hbox{\hbox to0pt
{\kern0.4\wd0\vrule height0.9\ht0\hss}\box0}}}}
\def\Cl{{\mathchoice
{\setbox0=\hbox{$\displaystyle\rm C$}\hbox{\hbox to0pt
{\kern0.4\wd0\vrule height0.9\ht0\hss}\box0}}
{\setbox0=\hbox{$\textstyle\rm C$}\hbox{\hbox to0pt
{\kern0.4\wd0\vrule height0.9\ht0\hss}\box0}}
{\setbox0=\hbox{$\scriptstyle\rm C$}\hbox{\hbox to0pt
{\kern0.4\wd0\vrule height0.9\ht0\hss}\box0}}
{\setbox0=\hbox{$\scriptscriptstyle\rm C$}\hbox{\hbox to0pt
{\kern0.4\wd0\vrule height0.9\ht0\hss}\box0}}}}
\def\Co{{\mathchoice
{\setbox0=\hbox{$\displaystyle\rm C$}\hbox{\hbox to0pt
{\kern0.4\wd0\vrule height0.9\ht0\hss}\box0}}
{\setbox0=\hbox{$\textstyle\rm C$}\hbox{\hbox to0pt
{\kern0.4\wd0\vrule height0.9\ht0\hss}\box0}}
{\setbox0=\hbox{$\scriptstyle\rm C$}\hbox{\hbox to0pt
{\kern0.4\wd0\vrule height0.9\ht0\hss}\box0}}
{\setbox0=\hbox{$\scriptscriptstyle\rm C$}\hbox{\hbox to0pt
{\kern0.4\wd0\vrule height0.9\ht0\hss}\box0}}}}
\def\Hl{{\mathchoice
{\setbox0=\hbox{$\displaystyle\rm H$}\hbox{\hbox to0pt
{\kern0.4\wd0\vrule height0.9\ht0\hss}\box0}}
{\setbox0=\hbox{$\textstyle\rm H$}\hbox{\hbox to0pt
{\kern0.4\wd0\vrule height0.9\ht0\hss}\box0}}
{\setbox0=\hbox{$\scriptstyle\rm H$}\hbox{\hbox to0pt
{\kern0.4\wd0\vrule height0.9\ht0\hss}\box0}}
{\setbox0=\hbox{$\scriptscriptstyle\rm H$}\hbox{\hbox to0pt
{\kern0.4\wd0\vrule height0.9\ht0\hss}\box0}}}}
\def\Ol{{\mathchoice
{\setbox0=\hbox{$\displaystyle\rm O$}\hbox{\hbox to0pt
{\kern0.4\wd0\vrule height0.9\ht0\hss}\box0}}
{\setbox0=\hbox{$\textstyle\rm O$}\hbox{\hbox to0pt
{\kern0.4\wd0\vrule height0.9\ht0\hss}\box0}}
{\setbox0=\hbox{$\scriptstyle\rm O$}\hbox{\hbox to0pt
{\kern0.4\wd0\vrule height0.9\ht0\hss}\box0}}
{\setbox0=\hbox{$\scriptscriptstyle\rm O$}\hbox{\hbox to0pt
{\kern0.4\wd0\vrule height0.9\ht0\hss}\box0}}}}
\DeclareMathOperator{\MC}{\boldsymbol{\mathsf{M}}}
\DeclareMathOperator{\MCO}{\boldsymbol{\widehat{\mathsf{M}}}}

\def\dprime{{\prime\prime}}

\title{{\sf Quantum Spin Dynamics VIII.}\\ {\sf The Master Constraint}} 
\author{{\sf
Thomas 
Thiemann}\thanks{thiemann@aei.mpg.de}\thanks{tthiemann@perimeterinstitute.ca} 
\\
{\sf MPI f. Gravitationsphysik, Albert-Einstein-Institut,} \\
{\sf Am M\"uhlenberg 1, 14476 Potsdam, Germany}\\
\\
{\sf and}\\
\\
{\sf Perimeter Institute for Theoretical Physics}\\
{\sf 31 Caroline Street North, Waterloo, Ontario N2L 2Y5, Canada}}
\date{{\small\sf Preprint AEI-2005-xxx}}
\date{}

\begin{document}




\maketitle



\begin{abstract}
{\sf Recently the Master Constraint Programme (MCP) for 
Loop Quantum Gravity (LQG) 
was launched
which replaces the infinite number of Hamiltonian 
constraints by a single Master constraint. The MCP is designed  
to overcome the complications associated with the non -- Lie -- algebra 
structure of the Dirac algebra of Hamiltonian constraints and was 
successfully tested in 
various field theory models. 

For the case of 3+1 gravity itself, so far only a positive quadratic form 
for 
the Master Constraint Operator was derived. In this paper we close this 
gap and prove that the quadratic form is closable and thus stems from a 
unique self -- adjoint Master Constraint Operator. The proof rests on 
a simple feature of the general pattern according to which Hamiltonian 
constraints in LQG are constructed and thus extends to arbitrary matter 
coupling and holds for any metric signature. 

With this result the existence of a physical Hilbert space for LQG is 
established by standard spectral analysis. }
\end{abstract}

\newpage



\section{Introduction}
\label{s1}

In the canonical approach to quantum gravity one is faced with the task to 
find a representation of the Dirac algebra $\mathfrak{D}$ of spatial 
diffeomorphism and Hamiltonian constraints. This algebra has the following
specific features:\\
\begin{itemize}
\item[1.] The Hamiltonian constraints are not spatially diffeomorphism 
invariant. In other words, the 
spatial diffeomorphism constraints form a subalgebra but no ideal
of $\mathfrak{D}$.
\item[2.]
The Poisson brackets of the Hamiltonian constraints are proportional to 
a spatial difeomorphism constraint, however, the coefficients of 
proportionality are non -- trivial functions on phase space, they are 
`structure functions' rather than structure constants. In other 
words, $\mathfrak{D}$ is not a Lie algebra in the strict sense of the 
word.
\end{itemize}
These features follow from very general properties of the hypersurface 
deformations of a foliation of spacetime and wil be ingredients of any 
canonical approach \cite{Hojmann}.

In order to quantise the constraints one has to write them 
in terms of (limits of) elements of a suitable subalgebra $\mathfrak{A}$ 
of the Poisson algebra of elementary functions on phase space which 
separates the points. In a second step one has to look for 
representations of $\mathfrak{A}$. Loop Quantum Gravity (LQG) 
(see \cite{LQG} for books and \cite{LQG1} for recent reviews) 
is nothing 
else than a canonical quantisation of General Relativity (GR) based on a 
specific choice of $\mathfrak{A}$, the so -- called holonomy -- flux 
algebra. These are functions on phase space which encode the magnetic and 
electric degrees of freedom of an SU(2) Yang -- Mills theory (plus matter)
following the real connection formulation of GR due to 
Ashtekar and Barbero \cite{Barbero}. Such functions are extremely natural 
from the 
point of view of (lattice) Yang -- Mills theory \cite{Gambini}.

Among the multitude of representations of $\mathfrak{A}$ one will be 
interested in those which are distinguished by physical selection 
criteria. One such criterion is a unitary representation of the spatial
diffeomorphism group (rather than a projective representation thereof).
Remarkably, it has been possible to show that such a representation is 
unique \cite{LOST}. More precisely, in general, any representation of a 
$^\ast-$algebra such as $\mathfrak{A}$ is a direct sum of cyclic 
representations and every cyclic representation comes from a state 
(positive linear functional) 
$\omega$ on $\mathfrak{A}$ via the GNS construction \cite{Bratteli}.
The elements $\varphi$ of the spatial diffeomorphism group act naturally 
on $\mathfrak{A}$ by  
automorphisms $a\mapsto \alpha_\varphi(a)$ which comes from pulling 
back the magnetic connection one form and electric field two form 
respectively. Hence, it suffices to look for invariant states (the 
associated representation of the diffeomorphism group is then 
automatically unitary) and the only such state is the Ashtekar -- Isham -- 
Lewandowski state \cite{AIL} $\omega_0$ which had been extensively used in 
LQG even 
before this uniqueness result transpired.

A particular feature of the AIL representation is the following:
\begin{itemize}
\item[3.] The spatial diffeomorphism group is represented unitarily but 
not weakly continuously. Hence, by Stone's theorem \cite{ReedSimon} 
the infinitesimal self -- adjoint generators, i.e. the Lie algebra of 
spatial
diffeomorphisms, does not exist in this representation.
\end{itemize}
Let us draw some simple conclusions from the results 1. -- 3.:
\begin{itemize}
\item[A.] Since the infinitesimal generators of spatial diffeomorphisms
appear in the classical algebra $\mathfrak{D}$ the representation 
$\omega_0$ of $\mathfrak{A}$ appears to be unsuitable to represent 
$\mathfrak{D}$.
\item[B.] A way out would be to exponentiate the spatial diffeomorphism 
and Hamiltonian constraints (`Weyl elements') and to deal with the 
associated groups they generate. However, while this is possible for 
the spatial diffeomorphism constraint (by construction of $\omega_0$), due 
to 2. there 
is no (Lie) group structure associated with the Hamiltonian constraints.
\item[C.] Another solution would be to solve first the spatial 
diffeomorphism constraint. Hence one would extract from the representation 
space ${\cal H}_0$ associated with $\omega_0$ the (generalised) spatially 
diffeomorphism invariant 
elements and assemble them into a Hilbert space ${\cal H}_{diff}$. Then 
the commutator 
between two Hamiltonian constraints would be trivial and obstacle 3. 
would be circumvented. While it is possible to construct ${\cal H}_{diff}$
\cite{ALMMT}, due to 1. ${\cal H}_{diff}$ does not carry a representation 
of the Hamiltonian constraints (they do not leave ${\cal H}_{diff}$ 
invariant).
\end{itemize}
Thus, 1. -- 3. seem to indicate that the structure of LQG must be changed:
The representations of the algebras $\mathfrak{D},\;\mathfrak{A}$ do not 
seem to be compatible. However, that is not necessarily the case: In 
\cite{QSD0,QSD1,QSD2,QSD3,QSD4,QSD5,QSD6,QSD7} the Hamiltonian constraints 
were quantised on ${\cal 
H}_0$ and 
their commutator indeed annihilates spatially diffeomorphsim invariant 
states (they are generalised zero eigenvectors). This is made possible 
because the right hand side of the Poisson bracket between two Hamiltonian 
constraints {\it is not a spatial diffeomorphism constraint}. It 
also involves 
the structure functions mentioned above and in quantum theory the 
associated composite operator (product of structure function and spatial 
diffeormorphism constraint) is less singular than the generator of spatial 
diffeomorphisms.

Yet, to answer the question whether the algebra of Hamiltonian constraints 
is properly 
implemented on ${\cal H}_0$ is currently unanswered. This is due to the 
fact that the Hamiltonian constraint is not a polynomial function
of the elementary variables and in order to mirror the classical Poisson 
bracket computation one has to have sufficient control over the spectrum 
of the volume operator \cite{RSVolume,ALVolume}. Work on this, using 
semiclassical 
techniques \cite{Semiclass}, is in progress. However, even after having 
resolved this issue, due to 2. so far it 
was not possible to find a constructive procedure to equip the 
physical states (generalised zero eigenvectors of the Hamiltonian 
constraint) with an inner product because group averaging techniques
(see e.g. \cite{ALMMT} and references therein) cannot cope with
constraint algebras with structure functions. It would 
therefore be more desirable to remove the tension between 
$\mathfrak{A},\;\mathfrak{D}$ from the outset and to replace one of them 
by a classically 
equivalent algebra such that there are common, manifest representations of 
both and such the physical Hilbert space can be constructed.

In \cite{Phoenix} the Master Constraint Progaremme was launched 
which proposes to replace $\mathfrak{D}$ by a much simpler Master 
Constraint Algebra $\mathfrak{M}$. Basically, the infinite number of 
Hamiltonian constraints are replaced by a single constraint, 
namely the weighted integral of their squares such that the associated 
Master Constraint $\MC$ is spatially diffeomorphism invariant. 
For this algebra, spatial 
diffeomorphisms form an ideal and the commutator of $\MC$ with itself is
trivial. One can show that $\mathfrak{D},\;\mathfrak{M}$ are classically
equivalent. The physical Hilbert is then readily available using standard 
spectral analysis techniques \cite{Phoenix,Test1} provided one manages to 
implement $\MC$ as a self -- adjoint operator $\MCO$ on 
either ${\cal H}_0$ or ${\cal H}_{diff}$ (and provided that the Hilbert
space is a direct sum of separable subspaces invariant for $\MCO$). 
To take the sum of squares of constraints
rather than the consraints themselves has successfully been tested for 
various  
toy models including those with an infinite number of degrees of freedom 
and with structure functions \cite{Test2}.

In \cite{Phoenix} we proposed a quadratic form $Q_{\MC}$ for the Master 
Constraint on a dense domain of ${\cal H}_{diff}$ ($Q_{\MC}$ is a graph 
changing, diffeomorphism invariant quadratic form and cannot exist
on ${\cal H}_0$, see \cite{Phoenix} for details). We 
also constructed 
a quadratic form $Q_{\MC_E}$ for the extended Master Constraint on ${\cal 
H}_0$ which also involves the weighted integral of the square of the 
spatial diffeomorphism constraint (possible because $Q_{\MC_E}$ is not 
graph 
changing). Two issues were left open in that paper:\\
1. A systematic derivation of $Q_{\MC}$ was not given.\\
2. It was not demonstrated that $Q_{\MC}$ is closable and is the quadratic 
form of a unique self -- adjoint operator $\MCO$. The same applies to 
$Q_{\MC_E}$. \\
\\
In this paper we close this gap:

In section two we derive $Q_{\MC}$ using the known regularisation of the 
Hamiltonian constraint. This is technically non -- trivial because 
the Hamiltonian constraint can only be defined on ${\cal H}_0$ while 
$Q_{\MC}$ can only be defined on ${\cal H}_{diff}$. The new technical tool
necessary for the derivation is the extension of the scalar product on 
${\cal H}_{diff}$ to all elements of the algebraic dual Cyl$^\ast$ of the 
space Cyl of finite linear combinations of spin network functions.

In section three we prove that $Q_{\MC}$ is closable. We also show that 
the proof extends to all matter coupling \cite{QSD5} and to $Q_{\MC_E}$.

In section four we display explicitly a separable subspace of ${\cal 
H}_{diff}$ which is left invariant by $\MCO$ and which should capture
the full physics of LQG.

In section five we show that the Master Equation, which is a condition 
on weak Dirac observables, is well defined without supplementing the 
Master Constraint with boundary terms in the presence of asymptotically 
flat boundary conditions.

In section five we conclude and outline the further steps in the task to 
solve the quantum dynamics of LQG.

\section{Derivation of the quadratic form of the master constraint}
\label{s2}

The derivation of the quadratic form of the master constraint will be 
given only for the full Lorentzian Hamiltonian 
constraint for pure gravity. We will show then that the same derivation 
applies to all matter coupling with just more terms to write.

\subsection{Strategy}
\label{s2.1}

The strategy to implement the Master constraint is as follows. Let 
${\cal T}(\epsilon)$ be a triangulation of $\sigma$ e.g. into 
tetrahedra 
$\Delta$ and 
denote by $\epsilon\to 0$ any limit in which the triangulation 
is infinitely refined subject to the constraints on a refinement that one 
uses in defining Riemannian integrals. 

Recall \cite{QSD0,QSD1} that up to a constant the   
Lorentzian Hamiltonian constraint of pure gravity is given by
\ba \label{2}
H(N) &=& \int_\sigma\; d^3x\; N(x) \;H(x)= a H_E(N)+b K(N)
\\
H_E(N) &=& \int_\sigma\; d^3x\; N(x)\; H_E(x)\;=\;\int_\sigma\; 
N(x) \; {\rm 
Tr}(F(x)\wedge \{A(x),V(R_x)\})
\nonumber\\
K(N) &=& \int_\sigma \; {\rm Tr}(
\{A(x),\{H_E(1),V(\sigma)\}\}
\wedge \{A(x),\{H_E(1),V(\sigma)\}\}\wedge
\{A(x),V(R_x)\})
\nonumber
\ea
where $x\mapsto H(x)$ denotes the Hamiltonian constraint
and 
$\sigma$ is a three manifold such that $\Rl\times \sigma$ is 
diffeomorphic to the spacetime manifold $M$.
Here $A$ is the gravitational SU(2) connection, $F$ its curvature, $N$
the lapse function, $R_x$ is any open region containing $x$ and 
\be \label{3}
V(R):=\int_R \; d^3x\; \sqrt{|\det(E)|}(x)
\ee
is the volume of $R$ with $E$ the electric field vector density. The 
non -- vanishing canonical brackets are 
$\{E^a_j(x),A_b^k(y)\}=\kappa\beta\delta^a_b \delta^k_j \delta(x,y)$
where $\kappa=8\pi G$, $G$ is Newton's constant and $\beta$ is the 
Immirzi parameter \cite{Immirzi}. The real constants $a,b$ in (\ref{2})
also depend on $\beta$. 

The integral
(\ref{2}) is the limit of the Riemann sum
\be \label{4}
H(N)=\lim_{\epsilon\to 0}\;\sum_{\Delta \in {\cal T}(\epsilon)}\; 
N(v(\Delta)) \; H(\Delta)
\ee
where $v(\Delta)$ is an interior point of $\Delta$ and
$H(\Delta)=H(\chi_\Delta)$ where $\chi_\Delta$ is the characteristic 
function of the set $\Delta$. That is, $H(\Delta)=H(N)_{N=\chi_\Delta}$.

Then the classical Master 
constraint as defined in \cite{Phoenix}
\be \label{1}
\MC\;=\;\int_\sigma\;d^3x\; \frac{[H(x)]^2}{\sqrt{\det(q)}(x)}
\ee
is likewise the limit of the Riemann sum 
\be \label{6.103}
\MC=\lim_{\epsilon\to 0}\;\sum_{\Delta \in {\cal T}(\epsilon)}\; 
\frac{H(\Delta)^2}{V(\Delta)}
\ee
where $H(\Delta)=H(\chi_\Delta),\;V(\Delta)=\int_\Delta d^3x 
\sqrt{|\det(E)|}$ as above.


We now choose w.l.g. the $R_x,\;x\in \Delta$ to actually coincide with 
$\Delta$ (only $x\in \partial\Delta$ are not interior points of $\Delta$ 
but these form a set of measure zero). Then 
\be \label{5}
C(\Delta):=\frac{H(\Delta)}{\sqrt{V(\Delta)}}=
\int_\Delta\; {\rm Tr}(F\wedge \frac{\{A,V(\Delta)\}}{\sqrt{V(\Delta)}})
=2\int_\Delta\; {\rm Tr}(F\wedge \{A,\sqrt{V(\Delta)}\})
\ee
where we used $\{.,V(\Delta)\}/\sqrt{V(\Delta)}=2\{.,\sqrt{V(\Delta)}\}$ 
and thus
\be \label{6.104}
\MC=\lim_{\epsilon\to 0}\;\sum_{\Delta \in {\cal T}(\epsilon)}\; 
C(\Delta)^2
\ee

Notice that $C(\Delta)$ is up to a factor of two the same as $H(\Delta)$ 
just that $V(\Delta)$ is replaced by $\sqrt{V(\Delta)}$. 
This is convenient because the $C(\Delta)$ can then be quantised 
precisely as the $H(\Delta)$ in \cite{QSD1,QSD5} with this simple change 
in the 
power of the 
volume operator. All the qualitative features remain the same, only the 
numerical values of the matrix elements of the corresponding regularised
operators 
$\hat{C}^\dagger_\epsilon(\Delta)$ change. Notice that in contrast to
$\hat{H}(\Delta)^\dagger$ the operator
$\hat{C}^\dagger_\epsilon(\Delta)$ depends in addition to the smearing 
function $\chi_\Delta$ on $\epsilon$ because we have to use 
$\sqrt{V(\Delta)}$ while for $\hat{H}(\Delta)^\dagger$ we may use 
$V(\sigma)$.  

We denote the 
quantisation of $C(\Delta)$, densely defined on the finite linear 
span $\cal D$ of spin network functions (which constitute a basis of 
${\cal 
H}_0$), by 
$\hat{C}^\dagger_\epsilon(\Delta)$ rather than 
$\hat{C}_\epsilon(\Delta)$ because of the definition of the dual operator
$\hat{C}'_\epsilon(\Delta)$ on the algebraic dual ${\cal D}^\ast$ 
(linear 
functionals on $\cal D$ without continuity assumptions): For 
$l\in{\cal D}^\ast$ 
and $f\in{\cal D}$ we have 
$[\hat{C}'_\epsilon(\Delta) l](f):=l(\hat{C}^\dagger_\epsilon(\Delta) f)$.
Classically $C(\Delta)$ is real valued so that 
$\hat{C}^\dagger_\epsilon(\Delta)$ qualifies as a quantisation of 
$C(\Delta)=\overline{C(\Delta)}$. Notice, however, that 
$\hat{C}^\dagger_\epsilon(\Delta)$ must not be symmetric for reasons of 
absence of anomalies in the constraint algebra, see e.g. \cite{Test1}
and references therein.\\
\\
One may therefore be tempted to simply compute the 
regularised dual operators $\hat{C}'_\epsilon(\Delta)$ on 
${\cal D}^\ast$ and then to restrict it to ${\cal D}^\ast_{diff}$ 
(the spatially diffeomorphism invariant elements of ${\cal D}^\ast$
\cite{ALMMT}).
Using the fact that $C(\Delta)$ is real valued we may write (\ref{6.104}) 
as 
\be \label{6.105}
\MC=\lim_{\epsilon\to 0}\;\sum_{\Delta \in {\cal T}(\epsilon)}\; 
\overline{C(\Delta)} C(\Delta) 
\ee
and since we must implement $\MCO$ directly on ${\cal H}_{Diff}$ 
(the Hilbert space completion of the finite linear span of spatially 
diffeomorphism group averaged spin networks functions \cite{ALMMT}) 
one would like to 
try to define the {\it quadratic form}
\be \label{6.106}
Q_{\MC}(l,l'):=
\lim_{\epsilon\to 0}\;\sum_{\Delta \in {\cal T}(\epsilon)}\; 
<l,(\hat{C}'_\epsilon(\Delta))^\ast\;
\hat{C}'_\epsilon(\Delta)\; l'>_{diff}
=\lim_{\epsilon\to 0}\;\sum_{\Delta \in {\cal T}(\epsilon)}\; 
<\hat{C}'_\epsilon(\Delta) \; l, 
\hat{C}'_\epsilon(\Delta) l'>_{diff}
\ee
where $(.)^\ast$ denotes the adjoint operation on ${\cal H}_{diff}$.
However, at least at finite $\epsilon$ equation (\ref{6.106}) is ill -- 
defined because we are using the scalar product on ${\cal H}_{diff}$
while $\hat{C}'_\epsilon(\Delta) l\not\in {\cal H}_{diff}$. 
In other words, just as the Hamiltonian constraint, $C(\Delta)$ is not 
spatially diffeomorphism invariant and $C'_\epsilon(\Delta)$ does not 
preserve ${\cal H}_{diff}$. For 
the same reason the adjoint operation, with respect to ${\cal H}_{diff}$ 
carried out in the second step is unjustified.\\

\subsection{New inner product on algebraic dual}
\label{s2.2} 

The hope is, of course, that (\ref{6.106}) makes sense in the limit 
$\epsilon\to 0$ when the corresponding classical quantity becomes 
spatially diffeomorphism invariant. The new tool to arrive at this and 
which we introduce here for the first time in LQG is 
to equip the space ${\cal D}^\ast$ with an inner product which reduces
to the one on ${\cal H}_{Diff}$ when evaluated on ${\cal 
D}^\ast_{diff}$.\\
\\
We will now, formally, define this inner product and start with
some preparations:\\ 
By ${\cal S}$ we denote the space 
of labels of spin network functions and we write $s$ for its elements
and $T_s$ for spin network functions. The orbit under (semianalytic 
\cite{LOST}) 
diffeomorphisms is given by $[s]:=\{\varphi\cdot s;\;\varphi\in {\rm 
Diff}(\sigma)\}$ where $s\mapsto\varphi\cdot s$ denotes the action of 
diffeomorphisms on spin network labels. Basically a spin network label
$s$ is a triple $s=(\gamma(s),j(s),I(s))$ consisting of a semianalytic 
graph $\gamma$, a labelling $j$ of its edges with non -- vanishing spin 
quantum numbers $j$ and a labelling of its vertices with gauge invariant
intertwiners $I$. Then $\varphi\cdot s=(\varphi(\gamma(s)),j(s),I(s))$.
Given a spin network 
diffeomorphism equivalence class $[s]$ we define the 
the non -- standard number or {\it Cantor aleph} 
\be \label{6.107}
\aleph([s]):=|[s]|:=|\{s'\in {\cal S};\;[s']=[s]\}|
\ee
as the size of the orbit $[s]$. 
Now recall \cite{ALMMT} that preferred elements of
${\cal D}^\ast_{diff}$ were given by
\be \label{6.108}
l_{[s]}:=\sum_{s'\in [s]} \;<T_{s'},.>_{kin},\;\eta(T_s)=
\eta_{[s]} l_{[s]}
\ee
with positive numbers $\eta_{[s]}$ and 
\be \label{6.109}
<\eta(T_s),\eta(T_{s'})>_{diff}=\eta(T_{s'})[T_s]
\ee
Here $\eta$ denotes the group averaging or rigging map introduced in 
\cite{ALMMT} and $<.,.>_{kin}$ denotes the inner product on ${\cal H}_0$.

An arbitrary element of ${\cal D}^\ast$ is of the form 
$l=\sum_{s\in {\cal S}}\; c_s\; <T_s,.>_{kin}$.
Formally, we may define an inner product $<.,>_\ast$ 
on ${\cal D}^\ast$ by
\ba \label{6.110}
<l,l'>_\ast &:=&
\sum_{s,s'} \overline{c_s} c'_{s'} <<T_s,.>_{kin},<T_{s'},.>_{kin}>_\ast
\nonumber\\
&:=&
\sum_{s,s'} \overline{c_s} c'_{s'} 
<T_{s'},T_s>_{kin} 
\frac{\sqrt{\eta_{[s]}\eta_{[s']}}}{\sqrt{\aleph([s])\aleph([s'])}}
=\sum_{s} \overline{c_s} c'_{s} \frac{\eta_{[s]}}{\aleph([s])}
\ea
This reproduces the inner product between the $\eta_{[s]}$ which 
correspond to $c_{s'}=\chi_{[s]}(s')$. It also formally corresponds to 
formally extending (\ref{6.110}) to ${\cal H}_{kin}$ with 
\be \label{6.111}
<T_s,T_{s'}>_\ast:=<T_s,T_{s'}>_{kin} 
\frac{\sqrt{\eta_{[s]}\eta_{[s']}}}{\sqrt{\aleph([s])\aleph([s'])}}
\ee
but of course elements of ${\cal H}_{Kin}$ have zero norm in this inner 
product. Hence by far not all elements of ${\cal D}^\ast$ are normalisable 
in this inner product and many elements have zero norm with respect to it.
By passing to the quotient by the null vectors and completing we may
turn the normalisable elements of ${\cal D}^\ast$ into a Hilbert 
space ${\cal H}_\ast\subset {\cal D}^\ast$. Notice 
that (\ref{6.110}) is the first inner product to be proposed on 
(a subset of) ${\cal D}^\ast$. 

It is curious to note that we may formally define a partial isometry 
\be \label{6.111a}
V:\;{\cal H}_\ast \to {\cal H}_{Kin};\; l=\sum_s\; c_s\; <T_s,.>_{Kin}
\mapsto \tilde{l}=\sum_s \; c_s\; 
\sqrt{\frac{\eta_{[s]}}{\aleph([s])}}\; T_s
\ee
so that we may formally identify $<.,.>_\ast$ with the kinematical inner 
product
$<.,.>_{kin}$ under the map $l\mapsto \tilde{l}$. 

In our application of $<.,.>_\ast$ quotients of non -- standard numbers 
will appear and this is a subtle issue in general \cite{Aleph}. 
Fortunately, the quotients we will find all equal unity or zero by 
inspection.

\subsection{Derivation of the quadratic form}
\label{s2.3}

The idea is then to use $<.,.>_\ast$ and its associated adjoint operation 
to define (\ref{6.106}) properly, that is,  
\be \label{6.112}
Q_{\MC}(l,l'):=
\lim_{\epsilon\to 0}\;\sum_{\Delta \in {\cal T}(\epsilon)}\; 
<l,(\hat{C}'_\epsilon(\Delta))^\ast\;
\hat{C}'_\epsilon(\Delta)\; l'>_\ast
=\lim_{\epsilon\to 0}\;\sum_{\Delta \in {\cal T}(\epsilon)}\; 
<\hat{C}'_\epsilon(\Delta)\; l,
\hat{C}'_\epsilon(\Delta) \;l'>_\ast
\ee
where $(.)^\ast$ is now the adjoint operation on ${\cal H}_\ast$ and 
(\ref{6.112})  
is now well -- defined. To evaluate $<.,.>_\ast$ we write
\be \label{6.113}
\hat{C}'_\epsilon(\Delta) l=\sum_{s\in{\cal S}}\;c^l_s(\Delta,\epsilon)
<T_s,.>_{kin} \;\;\Rightarrow \;\;
c^l_s(\Delta,\epsilon)=l(C^\dagger_\epsilon(\Delta) T_s)
\ee
Hence (\ref{6.113}) becomes
\ba \label{6.114}
Q_{\MC}(l,l')
&=& \lim_{\epsilon\to 0}\;\sum_{\Delta \in {\cal T}(\epsilon)}\; 
\sum_s\; \overline{c^l_s(\Delta,\epsilon)}\;
c^{l'}_s(\Delta,\epsilon)\;\frac{\eta_{[s]}}{\aleph([s])}
\nonumber\\
&=& \lim_{\epsilon\to 0}\;\sum_{\Delta \in {\cal T}(\epsilon)}\; 
\sum_{[s]}\; \frac{\eta_{[s]}}{\aleph([s])}\;\sum_{s'\in [s]}
\; \overline{c^l_{s'}(\Delta,\epsilon)}\;
c^{l'}_{s'}(\Delta,\epsilon)
\ea
We notice that for given $l,l'$ only a finite number of $[s]$ contribute 
to (\ref{6.114}): Namely, both $l,l'$ are finite linear combinations of 
the $l_{[s_1]}$ in (\ref{6.108}), hence it suffices to show that for any
$[s_1],[s_2]$ the numbers 
\be \label{6.115}
\overline{c^{l_{[s_1]}}_{s'}(\Delta,\epsilon)}\;
c^{l_{[s_2]}}_{s'}(\Delta,\epsilon)
\ee
are non -- vanishing only when $s'\in [s]$ and $[s]$ ranges over a 
finite number of classes. In order that 
$c^{l_{[s_1]}}_{s'}(\Delta,\epsilon)\not=0$ we must have that 
$\hat{C}^\dagger_\epsilon(\Delta)T_{s'}$ is 
a finite linear combination of spin network states which involves at 
least one of the $T_{s_1'}$ with $s_1'\in [s_1]$. But from the 
explicit action of $\hat{C}^\dagger_\epsilon(\Delta)$ \cite{QSD1} it is 
clear 
that for each $s_1'\in [s_1]$ there is only a finite set 
${\cal S}(s_1')$ of $s'$ with this 
property. Moreover, for each $s_1'\in [s_1]$ the number of elements 
of ${\cal S}(s_1')$ is the same and 
the classes of the elements of ${\cal S}(s_1')$ do not depend 
on the representative $s_1'\in [s_1]$. Denote the finite set of these 
classes by $[{\cal S}]([s_1])$. 

The sum over $[s]$ in (\ref{6.114}) is therefore only over the finite set
$[{\cal S}]([s_1])\cap [{\cal S}]([s_2])$ for 
$l=l_{[s_1]},\,l'=l_{[s_2]}$, hence for any $l,l'\in {\cal 
D}_{diff}\subset {\cal H}_{Diff}$ the 
sum over $[s]$ in (\ref{6.114}) is finite\footnote{${\cal D}_{diff}$ is 
the dense subset of ${\cal H}_{diff}$ consisting of the finite linear 
span of the $l_{[s]}$. Both are subspaces of ${\cal D}^\ast_{diff}$.}. We 
may therefore 
interchange the sum $\sum_{[s]}$ with the $\sum_{\Delta}$ and the limit
$\lim_{\epsilon\to 0}$ and arrive at 
\be \label{6.116}
Q_{\MC}(l,l')
=\sum_{[s]}\; \frac{\eta_{[s]}}{\aleph([s])}\;
\lim_{\epsilon\to 0}\;\sum_{\Delta \in {\cal T}(\epsilon)}\; 
\sum_{s'\in [s]}
\; \overline{c^l_{s'}(\Delta,\epsilon)}\;
c^{l'}_{s'}(\Delta,\epsilon)
\ee
Fix $s'\in [s]$ and consider $\hat{C}^\dagger_\epsilon(\Delta) T_{s'}$.
From \cite{QSD1} we know that this can be written in the form 
\be \label{6.117}
\hat{C}^\dagger_\epsilon(\Delta) T_{s'}=\sum_{v\in V(\gamma(s'))\cap 
\Delta} \;\hat{C}^\dagger_{\epsilon,\delta,\gamma(s'),v} T_{s'}.
\ee
where the operators $\hat{C}^\dagger_{\epsilon,\delta\gamma(s'),v}$ 
involve only 
degrees of 
freedom associated with edges of $\gamma(s')$ in the vicinity of $v$
and additional loops attached in a neighbourhood of $v$ 
which have to be chosen within 
the 
diffeomorphism invariance class specified in \cite{QSD1} and whose choice 
has been denoted by a choice function $\delta$.

For sufficiently small $\epsilon$ each $\Delta$ contains at most one 
vertex and the sum over $\Delta$ therefore reduces to the 
finite set ${\cal T}(\epsilon,s')$ of those $\Delta's$ containing 
precisely one vertex of $\gamma(s')$. We may therefore interchange 
the sum $\sum_{s'}$ with the $\sum_{\Delta}$ and the limit $\epsilon\to 
0$ and obtain
\ba \label{6.118}
Q_{\MC}(l,l')
&=&\sum_{[s]}\; \frac{\eta_{[s]}}{\aleph([s])}\;
\sum_{s'\in [s]}
\lim_{\epsilon\to 0}\;\sum_{\Delta \in {\cal T}(\epsilon,s')}\; 
\; \overline{c^l_{s'}(\Delta,\epsilon)}\;
c^{l'}_{s'}(\Delta,\epsilon)
\nonumber\\
&=&\sum_{[s]}\; \frac{\eta_{[s]}}{\aleph([s])}\;
\sum_{s'\in [s]}
\lim_{\epsilon\to 0}\;\sum_{v\in V(\gamma(s')}\;
\; \overline{c^l_{s'}(v,\epsilon,\delta)}\;
c^{l'}_{s'}(v,\epsilon,\delta)
\nonumber\\
&=&\sum_{[s]}\; \frac{\eta_{[s]}}{\aleph([s])}\;
\sum_{s'\in [s]} \;
\sum_{v\in V(\gamma(s'))}\;\; \overline{c^l_{s'}(v,\delta)}\;
c^{l'}_{s'}(v,\delta)
\nonumber\\
&=&\sum_{[s]}\; \frac{\eta_{[s]}}{\aleph([s])}\;
\sum_{s'\in [s]} \;
\sum_{v\in V(\gamma(s'))}\;\; \overline{c^l_{s'}(v)}\;
c^{l'}_{s'}(v)
\ea
where
\be \label{6.119}
c^l_{s'}(v,\epsilon,\delta)=l(\hat{C}^\dagger_{\epsilon,v,\delta} T_{s'})=
l(\hat{C}^\dagger_{v,\delta} T_{s'})=
l(\hat{C}^\dagger_{v,\delta_0} T_{s'})=
:c^l_{s'}(v)
\ee
Here in the second equality of (\ref{6.119}) the $\epsilon$ dependence 
coming from 
$V(\delta)$ has dropped out since $\Delta$ is so small that it contains
only $v\in V(\gamma(s'))$ and in the third equality we could fix 
$\delta=\delta_0$ by spatial diffeomorphism invariance of $l$.
In the second step in (\ref{6.118}) the sum over the contributing $\Delta$ 
could be 
replaced by sum the over vertices and since then nothing depends on 
$\epsilon$
any more the limit $\epsilon\to 0$ is trivial.

We now claim that 
\be \label{6.120}
a(s'):=\sum_{v\in V(\gamma(s'))}\;\; \overline{c^l_{s'}(v)}\;
c^{l'}_{s'}(v)
\ee
only depends on the class $[s]$ of $s'$. Indeed, 
\ba \label{6.121}
a(\varphi\cdot s') 
&=&
\sum_{v\in V(\gamma(\varphi\cdot s'))}\;\; 
\overline{c^l_{\varphi\cdot s'}(v)}\;
c^{l'}_{\varphi\cdot s'}(v)
\nonumber\\
&=& \sum_{v\in \varphi(V(\gamma(s')))}\;\; 
\overline{c^l_{\varphi\cdot s'}(v)}\;
c^{l'}_{\varphi\cdot s'}(v)
\nonumber\\
&=& \sum_{v\in V(\gamma(s'))}\;\; 
\overline{c^l_{\varphi\cdot s'}(\varphi(v))}\;
c^{l'}_{\varphi\cdot s'}(\varphi(v))
\ea
but 
\ba \label{6.121a}
c^l_{\varphi\cdot s'}(\varphi(v))
&=& l(\hat{C}^\dagger_{\varphi(v),\delta_0} \hat{U}(\varphi) T_{s'})
=l(\hat{U}(\varphi) \hat{C}^\dagger_{v,\delta'_0}  T_{s'})
\nonumber\\
&=& l(\hat{C}^\dagger_{v,\delta'_0}  T_{s'})
=l(\hat{C}^\dagger_{v,\delta_0}  T_{s'})
\nonumber\\
&=& c^l_{s'}(v)
\ea
where in the first step we used that Diff$^\omega_{sa}(\sigma)$ is 
unitarily implemented \cite{ALMMT}, in the second we have used the 
covariance relation up to a diffeomorphism 
$\hat{U}(\varphi) \hat{H}_{\delta_0}(N) \hat{U}(\varphi)^{-1}
=\hat{H}_{\delta_0'}(\varphi^\ast N)$  
established in \cite{QSD1} under which the choice
$\delta_0$ may change to some $\delta_0'$ but two choices are related by a 
diffeomorphism 
\cite{QSD1}
and in the last two steps we used diffeomorphism invariance of $l$.

It follows that all the $\aleph([s])$ terms in the sum $\sum_{s'\in [s]}$
are identical. Let $s_0([s])$ be a representative of $[s]$ then we may 
finish our derivation and get the final result
\be \label{6.122}
Q_{\MC}(l,l')
=\sum_{[s]}\; \eta_{[s]}\;
\sum_{v\in V(\gamma(s_0[s]))}\;\; 
\overline{l(\hat{C}^\dagger_v T_{s_0([s])})}\;
l'(\hat{C}^\dagger_v T_{s_0([s])})
\ee
We have dropped the irrelevant label $\delta_0$. Since we showed 
that the sum over $[s]$ collapses to a finite number of terms,
(\ref{6.122}) is well -- defined.\\
\\
Readers who dislike the formal steps performed involving division by and 
summing over $\aleph([s])$ terms may take (\ref{6.122}) as a definition.
Alternatively, one may dive into the field of non -- standard analysis 
\cite{Aleph} and regularise the sums over uncountably infinite number
of terms.

\section{The master constraint operator}
\label{s3}

Having constructed a qudratic form densely defined on the dense subspace
${\cal D}_{diff}\subset {\cal D}^\ast_{diff}$ of ${\cal H}_{diff}$ given 
by the finite linear span of elements of the form $l_{[s]}$ we want to 
show that $Q_{\MC}$ is associated to a unique self -- adjoint operator
$\MCO$. This is not trivial as the following reveals \cite{ReedSimon}.
\begin{Definition} \label{def6.1} ~~~~\\
i)\\
Let $T$ be a densely defined operator on a Hilbert space $\cal H$. 
$\Gamma(T):=\{(l,Tl);\;l\in D(T)\}\subset {\cal H}\times 
{\cal H}$ is called the graph of $T$. $T$ is called closed if 
the set $\Gamma(T)$ is closed with respect to the inner product 
$<(l_1,l_2),(l_1',l_2')>=<l_1,l_1'>+<l_2,l_2'>$. $T$ is called closable if 
it has a closed extension to $D(\bar{T})$, that is, $D(T)\subset 
D(\bar{T})$ and $\bar{T}_{|D(T)}=T$. The smallest closed extension is 
called the 
closure $\bar{T}$.\\ 
ii)\\
A quadratic form $Q$ on a Hilbert space 
${\cal H}$ is a sesqui -- linear form on $D(Q)\times D(Q)$ where $D(Q)$
is a dense form domain. A quadratic form is called semibounded 
provided that $Q(l,l)\ge -c ||l||^2$ for some $c\ge 0$ and positive if 
$c=0$. A semibounded quadratic form $Q$ is called closed provided that 
$D(Q)$ 
is complete in the norm $||l||_{+1}=\sqrt{Q(l,l)+c||l||^2}$. If $Q$
is closed and $D'(Q)\subset D(Q)$ is dense then $D'(Q)$ is called a 
form core.
\end{Definition}
\begin{Theorem} \label{th6.1} ~~~\\
i)\\
Let $T$ be a symmetric operator ($D(T)\subset D(T^\ast)$, 
$T^\ast_{|D(T)}=T$). Then $T$ is closable, however, its closure 
may not be self -- adjoint ($D(\bar{T})\not=D(\bar{T}^\dagger$).\\
ii)\\
Let $Q$ be a semi -- bounded quadratic form. Then $Q$ may not be closable,
but if it is and the closure is semi -- bounded, then $Q$ is the 
quadratic 
form of a unique self -- adjoint operator $T$ according to 
$Q(l,l')=<l,T l'>=:Q_T(l,l')$.\\
iii)\\
Let $T$ be a positive, symmetric operator. Then the corresponding 
positive quadratic form $Q_T$ has a positive closure
$\overline{Q_T}$. The uniqe positive operator $\tilde{T}$ 
corresponding to that closure via $Q_{\tilde{T}}=\overline{Q_T}$
is called the Friedrichs extension of $T$. It may extend the closure 
$\bar{T}$ of $T$ and is the only self -- adjoint extension which contains 
$D(\overline{Q_T})$.   
\end{Theorem}
In our case we can take as $D(Q_{\MC}):={\cal D}_{diff}$ the finite 
linear span of the $l_{[s]}$. Our $Q_{\MC}$ is manifestly positive
and sesqui -- linear.
It remains to show that it is closable. The problem that one might 
encounter is the following: The Hilbert space ${\cal H}_{diff}$ has the
orthonormal basis $T_{[s]}:=l_{[s]}/\sqrt{\eta_{[s]}}$ and we would like to 
define an operator $\MCO$ densely on $D(Q_{\MC})$ by
\be \label{6.123}
\MCO\;T_{[s_2]}:=\sum_{[s_1]}\;Q_{\MC}(T_{[s_1]},T_{[s]})\;T_{[s_1]}
\ee
However, the right hand side should be an element of ${\cal H}_{diff}$, 
that is
\be \label{6.124}
||\MCO\;T_{[s_2]}||^2:=\sum_{[s_1]}\;|Q_{\MC}(T_{[s_1]},T_{[s_2]})|^2\;<\infty
\ee
Hence there is a convergence issue to be resolved.
\begin{Theorem} \label{th6.2} ~~~~\\
i. The positive quadratic form $Q_{\MC}$ (\ref{6.122}) is closable and 
induces a unique, positive self-adjoint operator 
$\MCO$ on ${\cal H}_{diff}$.\\ 
ii. Moreover, the point zero is contained in the point spectrum of $\MCO$.  
\end{Theorem}
\begin{proof} of Theorem \ref{th6.2}:\\
i)\\
Since, given $[s_2]$ the `matrix element' $Q_{\MC}(T_{[s_1]},T_{[s_2]})$ 
is finite for every $[s_1],[s_2]$ in order to prove convergence of 
(\ref{6.124}) it will 
be sufficient to show that $Q_{\MC}(T_{[s_1]},T_{[s_2]})\not=0$ for at 
most a finite number of $[s_1]$ only.\\
1.\\
Let us fix $[s_1],[s_2]$ and consider 
the term corresponding to $[s]$ in (\ref{6.122}). In order that it does 
not vanish the expression
\be \label{6.125}
\sum_{v\in V(\gamma(s_0[s]))}\;\; 
\overline{T_{[s_1]}(\hat{C}^\dagger_v T_{s_0([s])})}\;
T_{[s_2]}(\hat{C}^\dagger_v T_{s_0([s])})
\ee
must be non -- zero. Hence the spin network decomposition of 
$\hat{C}^\dagger_v T_{s_0([s])}$ must contain a term diffeomorphic 
to $T_{s_1}$ and a term diffeomorphic to $T_{s_2}$ for at least
one $v\in V(\gamma(s_0([s])))$. Let us estimate the number of 
$[s]$ for which this is possible. The action of 
$\hat{C}^\dagger_v$ on $T_{s_0([s])}$ consists of two terms, 
corresponding to $H_E(N)$ and $K(N)$ in (\ref{2}) respectively (with the 
changed 
power of the volume operator):\\
First term:\\
The first term adds an arc in between any possible pair of edges 
with two possible orientations and 
changes the spin of the two correponding adjacent segments by $\pm 1/2$. 
Therefore it adds two more vertices.
Working at the gauge variant level (there are more gauge variant SNWF's
than invariant ones) this also changes the magnetic quantum numbers at
the end points of all three edges by $\pm 1/2$ which results in an 
additional factor of $4^3$ at most. 
Hence per vertex of valence $n(v)$ we get this way no more than
$4\cdot 2 \cdot 4^3 n(v) (n(v)-1)/2=4^4 n(v)(n(v)-1)$ new spin network 
states from the first term.\\
Second Term:\\ 
The second term is the square of the first term as 
far as the counting of new states is concerned. Hence we get 
$4^8 n(v)^2(n(v)-1)^2$ new spin network states from the second term
depending on two more acrcs and four more vertices.\\
Now in order that 
any of those is diffeomorphic to $T_{s_1}$ the graph $\gamma(s_0([s]))$
must have one or two edges less than $\gamma(s_1)$ and two or four 
vertices less than $\gamma(s_1)$. Moreover, the spins 
of the segments of edges adjacent to the arcs must differ by $\pm 1/2$ 
and the magnetic quantum numbers of arcs and edges must differ by $\pm 
1/2$. We conclude that if $N_1$ is the maximal valence of a vertex 
of $\gamma(s_1)$ then the number of $[s]$ that can contribute is bounded 
by
$4^8 N_1^4 |V(\gamma(s_1)|$ which depends only on $[s_1]$.
The same applies to $s_2$ of course. The actually contributing number of 
$[s]$ is certainly smaller than the maximum of 
$4^8 N_1^4 |V(\gamma(s_1)|, 4^8 N_2^4 |V(\gamma(s_2)|$.\\
2.\\
Let us now fix $[s_2]$ and let $[s_1]$ run. There are only 
$4^8 N_2^4 |V(\gamma(s_2)|$ classes $[s]$ which can contribute no matter
which $[s_1]$ we choose. By a similar argument, for each of those $[s]$ 
the number of $[s_1]$ 
which lead to a non -- vanishing contribution is bounded by 
$4^8 N^4 |V(\gamma(s_0([s]))|+2$ where $N$ is the maximal vertex valence of
$\gamma(s_0([s]))$. Since $N=N_2$ and 
$|V(\gamma(s_0([s])))|\le |V(\gamma(s_2))|$ we conclude that 
$Q_{\MC}(T_{[s_1]},T_{[s_2]})$ is non -- vanishing for at most
$4^{16} N_2^8 |V(\gamma(s_0([s_2])))|^2$ of the classes $[s_1]$.

We thus have shown that there is a positive symmetric operator $\MCO$ 
with dense domain ${\cal D}_{diff}$, the 
finite linear span of the $T_{[s]}$, defined by (\ref{6.124})
whose quadratic form coincides with $Q_{\MC}$ on the form domain 
$D(Q_{\MC})={\cal D}_{diff}$. Hence by theorem \ref{th6.1} iii) 
$Q_{\MC}$ has a positive closure and induces a unique self -- adjoint 
(Friedrichs) extension of $\MCO$ by theorem \ref{th6.1} which we denote by 
$\MCO$ as well.\\
ii)\\
Notice that the construction of the solutions of $\hat{H}'(N) l=0$ for all 
$N$ (which produces zero {\it eigenvectors}, i.e. normalisable elements of
${\cal H}_{diff}$)
displayed explicitly in \cite{QSD2} can be directly transcribed 
to the construction of solutions to $\MCO l=0$. Namely, $\MCO l=0$
implies $Q_{\MC}(l,l)=0$ which in turn enforces 
$l(\hat{C}^\dagger_v T_{s_0([s])})=0$ for all $[s]$ and all 
$v\in V(\gamma(s_0([s]))$. This is equivalent to 
$l(\hat{C}^\dagger(N) T_s)=0$ for all $s$ and all $N$ where 
$\hat{C}^\dagger(N)$ is defined identicaly as $\hat{H}^\dagger(N)$ just 
that one of the volume operators is replaced by two times its square root.
Thus, in particular $T_{[s]}$ where $s$ has no extraordinary edges 
\cite{QSD2} are normalisable solutions.
\end{proof}
Hence the Master constraint operator has a point kernel at least as 
rich as the 
Hamiltonian constraint. Moreover, it gives us additional flexibility 
in the following sense: In order to have a consistent constraint algebra 
the action of the Hamiltonian constraint had to be trivial at the vertices 
that it creates itself. However, the Master constraint does not have 
to satisfy any non -- trivial constraint algebra, hence this restriction 
can be relaxed to be less local. Whether such modifications 
lead to a sufficiently large semiclassical sector is, of course, not 
clear a priori and is subject to a detailed semiclassical analysis 
\cite{Semiclass}.\\
\\
We close this section with a remark on matter coupling and the extended 
Master Constraint:
\begin{itemize}
\item[1.] {\it Matter coupling}\\
The derivation and proof of closure of the quadratic form of the Master 
Constraint remains the same for all known matter \cite{QSD5} because the 
essential part of the derivation and proof respectively was that the 
attachment of the loop (arc) to a given graph follows diffeomorphism 
covariant rules. This was done univeraslly for matter and geometry in 
\cite{QSD5}.
\item[2.] {\it Extended Master Constraint}\\
In contrast to the Master Constraint considered in the previous two 
sections the extended Master Constraint also involves the spatial
diffeomorphism constraint (or even the Gauss constraint). Its classical 
expression is given by \cite{Phoenix} 
\ba \label{6.125a}
\MC_E &=& \int_\sigma \; d^3x\; \frac{H^2+q^{ab} H_a H_b}{\sqrt{\det(q)}} 
\nonumber\\
\MC_{EE} &=& \int_\sigma \; d^3x\; \frac{H^2+q^{ab} H_a H_b + H_j 
H_j}{\sqrt{\det(q)}} 
\ea
where $H,\;H_a,\;H_j$ denote Hamiltonian, spatial diffeomorphism and 
Gauss constraint respectively. Both constraints are spatially 
diffeomorphism invariant. However, $\MC_E$ allows us to implement both 
the Hamiltonian and the spatial diffeomorphism constraint on ${\cal 
H}_{kin}$ (and $\MC_{EE}$ also the Gauss constraint in addition) provided 
we implement the corresponding operators in a non -- graph changing 
fashion. In \cite{Phoenix} we showed how to do that using the notion of a 
minimal loop which is a loop (average) within the graph on which the 
constraint 
acts. It follows that instead of using dual operators we can directly work 
with operators on ${\cal H}_{kin}$ and their adjoints so that the 
construction of the quadratic form can be sidestepped. 
%
\end{itemize}

\section{Physical inner product and Dirac observables}
\label{s4}

Given the self -- adjoint Master constraint operator $\MCO$ of the 
previous section one would now like to use the machinery of the Direct 
Integral Decomposition reviewed in \cite{Test1} in order to define the 
physical 
Hilbert space. However, there is one additional obstacle: While 
the 
spectral theorem holds also in non -- separable Hilbert spaces, the direct 
integral decomposition can be performed only in the separable case.
However, ${\cal H}_{diff}$ is not 
separable unless, possibly, if we admit semianalytic homeomorphisms 
which remove the continuous moduli \cite{Moduli}
for vertices of valence five or higher. Now 
using homeomorphisms is forbidden because we must use the volume operator 
\cite{ALVolume} rather than \cite{RSVolume} as shown in 
\cite{Tina3,Tina4} which depends on a $C^{(1)}$ structure 
and 
which is absolutly crucial in order that $\MCO$ or $\hat{H}'(N)$ be even 
densely defined. Thus, the direct integral method seems not to be 
applicable.

Fortunately, ${\cal H}_{diff}$ can be decomposed as an uncountably 
infinite direct sum of separable Hilbert spaces as follows \cite{Phoenix}: 
\begin{Definition} \label{def6.2} ~~~~~~~~~\\
We say that two embedded graphs $\gamma_1,\gamma_2$ are $\theta$ -- 
equivalent (or homotopic up to the degeneracy type) provided that
that there exists a semianalytic diffeomorphism $\varphi \in 
$Diff$^\omega_{sa}(\sigma)$ such that:\\
1. $V(\varphi(\gamma_1))=V(\gamma_2)$\\
2. $\varphi(\gamma_1),\gamma_2$ are topologically equivalent, that is,
all vertices have the same connectivities with other vertices and 
edges are braided (knotted) and oriented the same way. Denote by $b:
E(\varphi(\gamma_1))\to E(\gamma_2)$ the corresponding bijection.\\
3. at each $v\in V(\gamma_2)$ and for each triple $e_1,e_2,e_3 \in 
E(\varphi(\gamma_1))$ of distinct edges the corresponding sign functions 
in coincide, that is, 
$\epsilon(e_1,e_2,e_3)=\epsilon(b(e_1),b(e_2),b(e_3))$, where 
$\epsilon(e_1,e_2,e_3)$ equals $+1,-1,0$ respectively if the tangents 
of the edges on their common starting point are right oriented, left 
oriented or co -- planar respectively.\\ 
\end{Definition}
Denote by $[\Gamma]$ the set of diffeomorphism equivalence classes 
$[\gamma]$ of graphs $\gamma\in \Gamma$ and by $(\Gamma)$ the set of 
$\theta-$equivalence classes $(\gamma)$ of graphs. Given $(\gamma)$,
let $\Theta'_{(\gamma)}$ be the set of moduli that are necessary 
to specify all the $[\gamma']$ with $(\gamma')=(\gamma)$. Hence any
element $[\gamma]\in [\Gamma]$ is now uniquely specified by a pair
$((\gamma),\theta')\in (\Gamma)\times \Theta'_{(\gamma)}$.
Let 
\be \label{6.126}
\Theta':=\times_{(\gamma)\in (\Gamma)}\; \Theta'_{(\gamma)}\;\;
\ni \;\theta=\{\theta'_{(\gamma)}\}_{(\gamma)\in (\Gamma)}
\ee
Then the direct sum of Hilbert spaces 
\be \label{6.127}
{\cal H}_{diff}=\oplus_{[\gamma]\in [\Gamma]} {\cal H}^{[\gamma]}_{diff}
\ee
where ${\cal H}^{[\gamma]}_{diff}$ is the closure of the finite linear 
span of $T_{[s]}$ with non -- trivial representations on all edges
can be decomposed also as 
\be \label{6.128}
{\cal H}_{diff}={\cal H}^0_{diff} \oplus
\oplus_{\theta'\in\Theta'}\;
\oplus_{(\gamma)\in (\Gamma)-(\Gamma)_0}\; 
{\cal H}^{((\gamma),\theta'_{(\gamma)})}_{diff}
=:\oplus_{\theta'\in\Theta'}\;{\cal H}^{\prime\theta'}_{diff}
\ee
where ${\cal H}^0_{diff}=\oplus_{[\gamma]\in [\Gamma]_0} {\cal 
H}_{[\gamma]}$ and $(\Gamma)_0=[\Gamma]_0$ is the subset of graphs 
without moduli.
We claim that all the ${\cal H}^{\prime\theta'}_{diff}$ are separable and 
mutually unitarily equivalent. Unitary equivalence is clear, we just have 
to map the corresponding points $\theta'$. Separability follows from the 
fact that at fixed $\theta$ a spin network label equivalence class is 
completely specified 
by 1. the number of vertices and their connectivities, 2. by the braiding 
and orientation of the corresponding edges and 3. by the spin and 
intertwining quantum numbers. Each of the three label sets is countable,
hence it has the cardinality of $\Nl^3$ which is countable.

Unfortunately, the sectors ${\cal H}^{\prime\theta'}_{diff}$ are 
generically not left 
invariant by $\MCO$:  This follows from the fact that $\MCO$ can have 
non -- vanishing matrix elements between $T_{[s]},T_{[s']}$ where 
$\gamma(s),\gamma(s')$ differ by an arc. Now $(\gamma(s)),(\gamma(s'))$
have the same moduli space, 
$\Theta'_{(\gamma(s))}=\Theta'_{(\gamma(s'))}$, because the two three -- 
valent vertices created by the arc do not require additional moduli
information and the $[s']$ obtained from $[s]$ is such that the modulis
coincide. However, in the Hilbert space 
${\cal H}^{\prime\theta}_{diff}$ the moduli assigned to $(\gamma(s'))$ 
might be different 
from those assigned to $(\gamma(s))$. Hence the $\theta$ sectors described 
above are not preserved. We could, of course, identify by hand these 
$\theta-$sectors 
and make ${\cal H}_{diff}$ separable altogether. The motivation for doing 
that is that every $\theta-$sector presumably already contains the 
physically 
relevant information encoded by $(\Gamma)$. But this is not what the 
formalism forces us to do.

It is therefore safer to do something else:
We can combine the $\theta-$moduli classification 
with the classification by sources ${\cal S}_0$ and derived spin
nets ${\cal S}_n(s_0)$ of level $n$ developed in \cite{QSD2}
as follows:\\ Denote by $[{\cal S}_0]$ the set of 
diffeomorphism equivalence classes of sources. For any
two representatives $s_1([s_0]),\;s_2([s_0])\in {\cal S}_0$ 
the set of diffeomorphism equivalence classes of the members of the 
derived 
spin nets of level $n$ 
of ${\cal S}_n(s_1([s_0])),\; {\cal S}_n(s_2([s_0]))$ coincide, i.e.
they depend only on $[s_0]$. We will denote this set therefore by 
$[{\cal S}_n]([s_0])$. We notice that the moduli parameters of all the 
$[s]\in [{\cal S}_n]([s_0]),\;n=0,1..$ are completely determined 
by those of $[s_0]$.
The completion of the finite linear span of these $T_{[s]}$ will be 
denoted ${\cal H}^{[s_0]}_{Diff}$ and this Hilbert space is separable by 
construction. Now the following issue arises: The action of $\MCO$ 
consists in adding and removing arcs to a graph and sometimes it reduces
the valence of a vertex by one or two units. It therefore happens 
that given $[s_0]\not=[s'_0]$ with $(s_0)=(s_0')$ that the set
$[{\cal S}_n]([s_0])\cap[{\cal S}_n]([s'_0])$ is not empty. For instance a 
five -- valent vertex, which has moduli, could be turned into a three -- 
valent one which does not have moduli. Hence it is almost but not quite 
true that ${\cal H}_{diff}$ is the uncountable 
direct sum of the ${\cal H}^{[s_0]}_{diff},\;[s_0]\in[{\cal S}_0]$. 

Let us write $[s_0]=((s_0),\theta_{(s_0)}:=\theta_{(\gamma(s_0))})$ where 
$(s_0)$ is the 
$\theta-$equivalence class of $s_0$ which is determined by the 
$(\gamma(s_0))$. Let $({\cal S}_0)$ be the set of those $(s_0)$ and let
$\Theta'$ be the collection of the $\theta_{(s_0)},\;(s_0)\in 
({\cal S}_0)$. Then 
\ba \label{6.129}
{\cal H}_{diff}&=&
\cup_{\theta\in \Theta}\; 
{\cal H}^{\theta}_{diff}:=
\cup_{(s_0)\in ({\cal S}_0)}\;
{\cal H}^{((s_0),\theta_{(s_0)})}_{diff}
\ea
Notice that the unions are almost direct sums but not quite as just 
pointed out. However, each of the spaces  
${\cal H}^{\theta}_{Diff}$ is a separable 
and $\MCO-$invariant subspace of ${\cal H}_{Diff}$ and all of them 
are mutually isomorphic.
Moreover, each of them contains information about all $\theta-$equivalence 
classes of spin network states and therefore
all the physically relevant information.\\
\\
Thus, while these are not sectors in the strict sense, we may just 
pick one of these subspaces and 
apply the direct integral decomposition method to it.
\begin{Theorem} \label{th6.3} ~~~~\\
There is a unitary operator $V$ such that 
$V{\cal H}^\theta_{diff}$ is the direct integral Hilbert space
\be \label{6.131}
{\cal H}^\theta_{diff}\propto \int_{\Rl^+}^\oplus \;
d\mu(\lambda)\;{\cal H}^\theta_{diff}(\lambda)
\ee
where the measure class of $\mu$ and the Hilbert spaces 
${\cal H}^\theta_{Diff}(\lambda)$, in which $V\MCO V^{-1}$ acts 
by multiplication by $\lambda$, are $\mu-$uniquely determined. 

The physical Hilbert space is given by\footnote{The spaces 
${\cal H}^\theta_{diff}(\lambda)$ are defined up to measure $\mu$ zero 
sets. See \cite{Test1} for physical criteria to choose an appropriate 
candidate.}  
${\cal H}^\theta_{phys}={\cal H}^\theta_{diff}(0)$.
\end{Theorem}
Dirac observables could now be constructed from spatially diffeomorphism
invariant operators which preserve any ${\cal H}_{diff}^\theta$ e.g. by 
using 
the ergodic projection technique of \cite{Phoenix} or by the partial 
observable technique of \cite{Bianca}. Any spatially
diffeomorphism invariant operator regularised in the same fashion as the 
Hamiltonian constraint operator has the property to preserve 
each of the subspaces ${\cal H}_{diff}^\theta$ separately, hence this is 
no restriction.

\section{The classical master equation for selecting weak Dirac 
observables}
\label{s5}

In the case of boundaries of $\sigma$, the classical Hamiltonian 
constraint has to be supplemented by boundary terms in order to be 
functionally differentiable (i.e. its Hamiltonian vector field is well 
defined) for lapse functions which do not vanish on the boundaries. 
At first sight, the Master Constraint needs to be twice functionally 
differentiable in order that the {\it Master Equation} definition of weak 
Dirac observables
$\{O,\{O,\MC\}\}_{\MC=0}=0$ \cite{Phoenix} makes sense and since there are 
no smearing 
(lapse) functions involved the issue of boundary terms could be non -- 
trivial.
We will now show that the Master constraint is actually more regular than 
the Hamiltonian constraint as far as the Master Equation is concerned and 
the issue of boundarty terms does not 
arise in the case of asymptotic flatness (no interior boundaries). We will 
just sketch this for the piece $H_E(x)$ and for pure 
gravity, similar 
remarks hold for the full Lorentzian constraint and 
matter coupling.\\
\\
Let us write $\MC=\int \; d^3x\; 
C(x)^2,\;C(x)^2=H_E(x)^2/\sqrt{\det(q)}(x)$. 
Recall \cite{Boundary}
that the boundary conditions at spatial infinity are such that the 
components of 
$A$ fall off as $r^{-2}$ while the components of $E-E_0$ fall off as 
$r^{-1}$ where $E_0$ is the fixed boundary value of $E$ compatible with 
the 
Euclidean metric. Here $r$ is with respect to an asymptotically Cartesian 
system of coordinates and with respect to asymptotic reflections 
$A$ and $E-E_0$ respectively have to be odd and even respectively.
It follows that the tangent vectors $\delta A,\;\delta E$ fall off
as $A,\; E-E_0$ and have the same reflection properties.

We conclude from the definition of $H_E$ in (\ref{2}) that $C(x)$
falls off as $r^{-3}$ so that the integral of the Master constraint itself
converges. In what follows we will symbolically write $df,Df$ for 
a partial or SU(2) covariant
derivative of a function of $f$ which will be enough to determine the 
fall off properties of various terms. We will also write $Q$ for a generic 
funcion of $E$ alone which asymptotically convereges to a constant.
We assume of course that $O$ itself 
is functionally differentiable, that is
\be \label{7}
\delta O=\int_\sigma [I\cdot \delta A+J\cdot \delta E]
\ee
converges, so that $I$ falls off at least as $r^{-1}$ with even parity and 
$J$ as $r^{-2}$ with odd parity. The variation of $\MC$ itself gives
\be \label{8}
\delta\MC=\int_\sigma[(D (Q_1 C))\cdot \delta A+(Q_2 C)\cdot \delta 
E]
\ee
The first term involves an integration by parts but since $C$ falls off
as $r^{-3}$ no boundary term is picked up.

Combining (\ref{7}) and (\ref{8}) we find for the first order Poisson 
bracket
\be \label{9}
\{O,\MC\}=\int_\sigma [I\cdot (Q_2 C)-J\cdot D(Q_1 C)]
\ee
The integrand falls off at least as $r^{-4}$ ands thus converges. Varying
(\ref{9}) again at $\MC=0$ we just need to consider the terms that result
from variations of $C$ (otherwise, that is when considering variations 
off the constraint surface $\MC=0$, we 
must make suitable assumptions about 
the variations of $I,J$). Hence
\be \label{10}
\delta\{O,\MC\}_{\MC=0}= 
\int_\sigma [I\cdot (Q_2 \delta C)-J\cdot D(Q_1 \delta C)]
=
\int_\sigma [I Q_2+(D J) Q_1] \cdot  \delta C
\ee
In the second step we had to perform an integration by parts in the second 
term. Since $\delta C$ falls off at least as $r^{-3}$ while $J$ falls off 
at least as $r^{-2}$ no boundary term is picked up. Performing a further 
integration by parts we can finish (\ref{10}) with the result
\be \label{11}
\delta\{O,\MC\}_{\MC=0}= 
\int_\sigma [(D (Q_1(I Q_2+(D J) Q_1))) \cdot  \delta A
+(F Q_4)(I Q_2+(D J) Q_1)\cdot \delta E]
\ee
where as before $F$ denotes the curvature of $A$. We have dropped a term 
proportional to $C$ in the variation of $E$. The intergral (\ref{11})
evidently converges because the integrand falls off at least as $r^{-4}$.

Combining (\ref{7}) and (\ref{11}) we find 
\be \label{12}
\{O,\{O,\MC\}\}_{\MC=0}= 
\int_\sigma [I\cdot (F Q_4)(I Q_2+(D J) Q_1)]
-J\cdot (D (Q_1(I Q_2+(D J) Q_1)))]
\ee
and the integral still converges since the integrand falls off as $r^{-4}$ 
at least. Notice that the parity properties never had to be used.\\
\\
We conclude that the Master Equation is well defined 
without additional boundary terms for the Master Constraint with 
asymptotically flat
boundary conditions at least on the constraint surface $\MC=0$ for once
functionally differentiable $O$ and off the constraint surface for 
suitable twice differentiable $O$.

\section{Conclusions}
\label{s6}

The results of \cite{Phoenix} and this paper establish that a self -- 
adjoint, positive Master Constraint Operator for LQG exists. The results 
of \cite{Test1} show that the existence of a physical inner product 
by direct integral methods is automatic. The results of \cite{Test2}
demonstrate that the Master Constraint Programme leads to the expected 
physical results in 
a large number 
of rather generic model situations, e.g., in examples with second class 
constraints, with structure functions, an infinite number of degrees of 
freedom etc.

Taken together, the Master Constraint Programme has good chances to 
overcome the difficulties that have hindered progress with the Hamiltonian 
constraint over the past decade. The next step is to check whether 
the Master Constraint has the correct classical limit 
and to develop approximation methods that enable to construct physical 
states and the physical inner product explicitly.\\
\\
\\
{\large Acknowledgements}\\
\\
We would like to thank Yongge Ma and Muxin Han, who have arrived at 
similar results from a different perspective, for discussions. 
It is a pleasure to thank Yongge
Ma for his generous hospitality during the author's visit to Bejing Normal 
University.
This research project has been supported in part by funds from NSERC of 
Canada to the Perimeter Institute for Theoretical Physics.

\end{document}